\documentclass[12pt,preprint]{aastex}

\def\gsim{\ \raise 3pt \hbox{$>$} \kern -8.5pt \raise -2pt \hbox{$\sim$}\ }
\def\lsim{\ \raise 3pt \hbox{$<$} \kern -8.5pt \raise -2pt \hbox{$\sim$}\ }

\begin{document}

\title{Millisecond microwave spikes: statistical study and application for
plasma diagnostics}

\author{I.~V.~Rozhansky\altaffilmark{1,2}, G.~D.~Fleishman\altaffilmark{3,2}, G.-L.~Huang\altaffilmark{1}}
%\author{Gregory D. Fleishman\altaffilmark{1,2}}
\altaffiltext{1}{Purple Mountain Observatory, National Astronomical
 Observatories, Nanjing 210008, P.R. China}
\altaffiltext{2}{Ioffe Institute for Physics and Technology, 194021
St.Petersburg,
 Russia}
\altaffiltext{3}{New Jersey Institute of Technology, Newark, NJ
07102}

\begin{abstract}
{We analyze a dense cluster of solar radio spikes registered at ~
4.5 -- 6 GHz by the Purple Mountain Observatory spectrometer
(Nanjing, China) operating in the 4.5 -- 7.5 GHz range with the 5 ms
temporal resolution. To handle with the data from the spectrometer
we developed a new technique utilizing a nonlinear multi-Gaussian
spectral fit based on "chi-squared" criteria to extract individual
spikes from the originally recorded spectra. Applying this method to
the experimental raw data we eventually identified about 3000 spikes
for this event, which allows for a detailed statistical analysis.
Various statistical characteristics of the spikes have been
evaluated, including intensity distributions, spectral bandwidth
distributions, and distribution of the spike mean frequencies. The
most striking finding of this analysis is distributions of the spike
bandwidth, which are remarkably asymmetric. To reveal the
underlaying  microphysics we explore the local trap model with the
renormalized theory of spectral profile of the electron cyclotron
maser (ECM) emission peak in a source with random magnetic
irregularities. The distribution of the solar spikes relative
bandwidth calculated within the local trap model represents an
excellent fit to the experimental data. Accordingly, the developed
technique may offer a new tool of studying very low levels of the
magnetic turbulence in the spike sources, when the ECM mechanism of
the spike cluster is confirmed.}
\end{abstract}


\section{Introduction}

Solar radio spikes are known to be very narrowband kind of solar
radio emission \citep{Benz1985,StaehliMagun1986} displaying a
typical bandwidth of the order of $1\%$. A more detailed study of
the bandwidth distribution and its correlations with other observed
parameters was performed by \cite{CsillaghyBenz1993}. They noted
that the bandwidth changes significantly from one event to another,
so no clear correlation between the bandwidth and the observing
frequency is visible. Thus, the bandwidth is more characteristics of
the event rather than function of frequency.

Then, no unique correlation between the bandwidth and radio flux of
the spikes was found: there were uncorrelated cases, as well as
correlated and anti-correlated cases. Based on  these results,
\cite{CsillaghyBenz1993} concluded that the observed bandwidth is
formed mainly by source inhomogeneity rather than natural bandwidth
of the underlying emission process.

\cite{MessmerBenz2000} applied a more advanced approach to determine
a minimum bandwidth of solar radio spikes, assuming that the minimum
bandwidth may correspond  to the natural bandwidth of the emission
process, while broader spikes are signature of the source
inhomogeneity. The minimum bandwidth for two considered events was
$0.17\%$ and $0.41\%$ respectively, implying the natural bandwidth
of the emission process to be less than the measured values.

The currently available data about the radio spikes appearing at the
main flare phase are pretty consistent with the idea that the source
of spike cluster is a loop filled by fast electrons and relatively
tenuous background plasma \citep{FlGaryNita2003}. The trapped fast
electrons have a nonthermal (power-law) energy spectrum and a
loss-cone angular distribution. Each single spike is generated in a
local source inside this loop when the local anisotropy is increased
compared with the averaged one to produce electron cyclotron maser
(ECM) emission. The assumed fluctuations of the pitch-angle
distribution of fast electrons can be produced by the magnetic
turbulence like in the turbulent model proposed by
\citep{BaKar2001}.

ECM emission is believed to be responsible for many kinds of solar,
planetary and stellar radiation
\citep{Stepanov1978,WuLee79,Holmanetal80,MelroseDulk1982,
SharVla1984,Wu85,Wingetal1988,
AschwBenz1988,Aschw1990,Barrowetal1994,
Stupp2000,Vlasov2002,LaBelleTreumann2002, Treumann_2006}. Recently
\citep{FlMel1998,FlGaryNita2003}, new important evidence for solar
radio spikes to be produced by ECM emission has been obtained.
However, although ECM emission fits well to many spike properties,
no direct comparison between the spike spectral properties and  the
ECM spectral properties has been made yet.

\cite{Hewittetal1982} suggested a simple kinematic estimate for the
ECM bandwidth
\begin{equation}\label{MD_band}
  \Delta \omega/\omega \sim (v/c)^2,
\end{equation}
where $\omega$ is the central frequency of the emission line,
$\Delta \omega$ is the spectral bandwidth, $v$ is a characteristic
velocity of fast particles responsible for the ECM generation. This
estimate has evidently a limited applicability region, since it does
not depend on fast electron pitch-angle distribution, viewing angle
etc. Moreover, it is not clear what value of $v$ should be used if
the energy distribution of fast electrons is rather broad like in
case of power-law spectra typical for solar flares.

The problem of ECM bandwidth attracted a lot attention in connection
with the fine structure of the terrestrial auroral kilometric
radiation (AKR) \citep{GurnettAnderson1981,
BaumbackCalvert1987,YoonWeatherwax1998,Pritchettetal1999}. Indeed,
the bandwidth of individual AKR peaks is observed to be as
small as $\Delta \omega/\omega \sim 10^{-3}$ %\\
\citep{GurnettAnderson1981} with the extreme values down to
$10^{-5}$ \citep{BaumbackCalvert1987}. It was shown recently
\citep{YoonWeatherwax1998} that the choice of a realistic
distribution function of the superthermal electrons may easily
provide the ECM bandwidth of the order of $10^{-3}$.

However, the results obtained for the auroral region cannot be
directly applied to the solar case because of important differences
in the source conditions. First of all, the plasma frequency to
gyrofrequency ratio
\begin{equation}\label{Y}
  Y=\omega_{pe}/\omega_{Be}
\end{equation}
is much less than unity for the AKR source, while is of the order of
unity or larger for the solar corona. Therefore, the corrections to
the wave dispersion provided by superthermal electrons may be
important for the AKR source \citep{YoonWeatherwax1998}, while are
typically negligible for the solar case. Then, the fundamental
extraordinary wave-mode is the most important if $Y \ll 1$, while
different wave-modes (fundamental or harmonic ordinary and harmonic
extraordinary) become important if $Y \sim 1$. The distributions of
the fast electrons are also believed to differ significantly for
these two cases: keV electrons are responsible for the AKR, while
broad distributions covered the range $10-10^3$keV at least arise
typically in solar flares.

Thus, the study of the ECM spectral properties for the conditions
typical for solar flares is largely an independent problem that
deserves particular attention and careful theoretical consideration.
So far, the natural bandwidth of the ECM emission in the standard
coronal conditions has been studied in detail by \cite{Fl_2004_AL}
who demonstrated that the relative ECM bandwidth typically belongs
to the range
\begin{equation}\label{natural_bandwidth}
  \Delta \omega/\omega \sim 0.1-0.4 \%,
\end{equation}
which is more than one order of magnitude less than  intuitive
estimate (\ref{MD_band}).

The effect of source inhomogeneity on the bandwidth of ECM peaks was
studied by \cite{PlatonovFleishman2001} in the linear approximation,
i.e., when the quasilinear saturation and nonlinear wave-wave
interactions are not important. The corresponding broadening related
to gradual non-uniformity of the solar corona was found to be rather
small, while the effect of \emph{random} inhomogeneities of the
magnetic field is typically important. \cite{Fl_2004b} developed a
renormalized theory of the ECM emission in the source with random
inhomogeneities of the magnetic field and found that relatively weak
magnetic inhomogeneities can provide strong  broadening of the ECM
peaks.

%However, the natural bandwidth of ECM emission and its dependence on
%various parameters was not studied in detail yet {\it for the solar
%case}, so nothing is known currently except the estimate
%(\ref{MD_band}) about the natural bandwidth.
%
%The properties of ECM emission are sensitive to the distribution of
%the fast electrons over momentum and pitch-angle. Recently
%\citep{Fl2003a,Fl2003b} it has been shown that the most appropriate
%model pitch-angle distribution is the gaussian loss-cone function:
%\begin{equation}
%\label{distr_mu_gauss} f_2(\mu) \propto \Biggl\{ \begin{array}{cc}
%\exp\left(- \left({\mu-\cos \theta_c \over \Delta \mu}\right)^2
%\right),&
%\mu > \cos \theta_c, \\
%1, & -\cos \theta_c < \mu < \cos \theta_c\\
%\exp\left(- \left({\mu +\cos \theta_c \over \Delta \mu}\right)^2
%\right),& \mu < -\cos \theta_c,
%\end{array}
%\end{equation}
%where $\mu=\cos \theta$ is the cosine of the electron pitch-angle,
%$\theta_c$ is the loss-cone angle, $\Delta \mu$ is the width of the
%loss-cone function. The distribution of the fast electrons over
%momentum is evidently a kind of power-law.

In this paper we analyze a dense cluster of solar radio spikes
registered at high frequencies above 4.5 GHz.  A special numerical
technique was developed to decompose partly overlapping spikes in
the cluster into individual gaussian spikes and yield large
statistically significant series of the spikes.
%. The proposed procedure yields large statistically significant
%series of the spikes sufficient for detailed statistical analysis of
%the spike
%properties. % with the gaussian spectral shape even in the case of
%overlapping spikes.
Applying this method to the experimental raw data we eventually
identified about 3000 spikes for the observed event, which allows
for a detailed statistical analysis.

The most striking finding of this analysis is distributions of the
spike bandwidth, which are remarkably asymmetric. The overall
bandwidth distribution has a characteristic skew shape with rapid
increase at low values of the relative bandwidth followed by maximum
at 0.6 \% and smooth tail approaching zero at approximately 3\%.  In
order to account for the essential features of the bandwidth
distributions we explicitly use the renormalized theory of spectral
profile of the ECM emission peak. The theory accurately takes into
account the fluctuations of the magnetic field in the spike source.
The bandwidth distribution obtained by the proposed theory is found
to agree excellently with the observed spike bandwidth distribution.

\section{Observations}

We analyze a dense cluster of solar radio spikes registered from
05:18:03-05:18:09 UT on April, 10, 2001, at 4.5 -- 6 GHz by the
Purple Mountain Observatory spectrometer (Nanjing, China) operating
in the 4.5 -- 7.5 GHz range with the temporal resolution of 5 ms.
The cluster occurred during a X2.3 flare on April 10, 2001
\citep{Asai_etal_2003, Chernov_etal_2006}, NOAA region 9415, located
close to the center of the solar disk (S23W06-08). The flare was
associated with a halo CME, meter-wavelength types II and IV bursts,
and strong microwave continuum burst.

\begin{figure}
%\leavevmode
 %\centering\epsfxsize=470pt \epsfbox[60 430 590 800]
  \epsscale{0.8}   \plotone{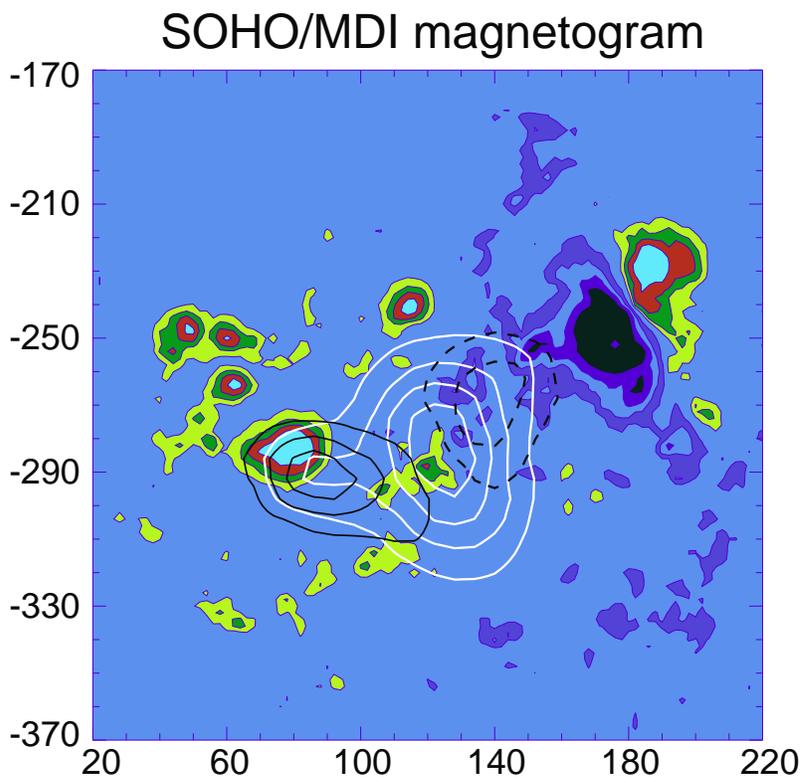} %{mdi_17GHz.ps}
\caption{ %C:\Radio_Data\Spike_Locations\
Context data at the time of the spike cluster. Color image: SOHO/MDI
magnetogram; the contour levels are selected to highlight  the
photospheric regions where the fundamental gyroharmonics (+1500 to
+2000 G, red, and -1500 to -2000 G, violet) and the pair of the
second and third harmonics (+500 to +1000 G, yellow, and -500 to
-1000 G, dark blue) fall into the 4.5-5.6 GHz spectral range. White
contours display the Stocks I distribution at the levels of
(0.2,~0.4,~0.6,~and~0.8) of $I_{max}$. Solid black contours display
the positive  Stocks V distribution (RCP emission) at the levels of
(0.05,~0.1,~and~0.15) of $I_{max}$. Dashed black contours display
the negative  Stocks V distribution (LCP emission) at the levels of
(-0.05,~and~-0.1) of $I_{max}$.
\label{fig:MDI} }
\end{figure}

The spike cluster occurred during a local impulsive peak of strong
long microwave burst with the absolute peak value in excess of 6000
sfu around 9.4 GHz. Highly polarized coherent emissions (LCP at 2
GHz and RCP at 3.75 GHz) were recorded by the Nobeyama Polarimeters
around the time of the spike cluster. Context data on the
photospheric magnetic fields and microwave emission at 17 GHz at the
time of spike cluster are shown in Figure~\ref{fig:MDI}.

\subsection{Instrumentation}

The data we use here are collected by the radio spectrometer at PMO,
China \citep{Xu_etal_2003}. It has 300 frequency channels per 3 GHz
band of 4.5-7.5 GHz with spectral resolution of 10 MHz and time
resolution of 5 ms, which observes daily between 1:00 and 9:00 UT.

The dynamic spectrum measured for the event studied is given in
Figure~\ref{fig:spectra}(a). Two most intensive regions of the spike
cluster can be seen around the second and the third seconds of the
given dynamic spectrum. As seen from the figure the spikes are not
well separated in the frequency domain but rather substantially
overlap producing a continuously fluctuating spectrum.

\subsection{Spike Resolution and Identification}

If the resolution of the instrument is high enough in both temporal
and spectral domains, each spike represents a 2D object in the
dynamic spectrum, which can, in particular, be characterized by
duration, bandwidth, and spectral drift \citep{Dabrowski_etal_2005}.
In many cases, however, the resolution in either spectral or
temporal domain is insufficient to fully resolve each spike. Let us
consider first the available temporal resolution against the spike
duration.

\begin{figure}[ht]
% \leavevmode
 %\centering\epsfxsize=200pt \epsfbox[65 400 630 800]
  \epsscale{0.5} \plotone{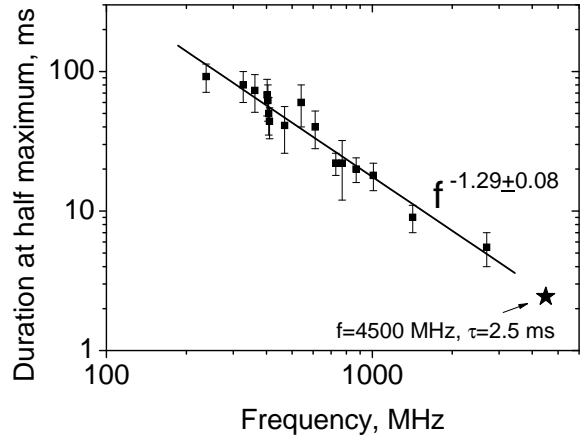} %{Durationfrequency.eps}
\caption{Observed duration of spikes at half maximum vs frequency of
observation based on published data: 237, 327, 408 and 610 MHz from
\citep{ZloKar1998} and P.Zlobec, private communication, 540 MHz from
\citep{Lipatovetal2002}, 362, 468, 730, 770, 870, 1010 MHz from
\citep{GuBenz1990}, and 1420 and 2695 MHz from
\citep{KarZl2002,KarZl2003}. The solid line is the best power-law
fit with the index $-1.29 \pm 0.08$. \label{fig:Durationfrequency}}
\end{figure}

Lower-frequency observations performed mainly at the decimetric
spectral range show that the spike duration is largely a function of
the emission frequency with rather weak scatter around the
regression curve \citep{GuBenz1990,KarZl2002,KarZl2003}, see also
Figure~\ref{fig:Durationfrequency} gathering all currently available
measurements of spike duration. The regression law found from this
% \IR corrected few misprints here - law (was 'low')
figure for the entire available spectral range, $237-2695$~MHz,
\begin{equation}
\label{Dur_obs} \tau \propto f^{-1.29 \pm 0.08},
\end{equation}
represents a corrected G\"udel-Benz law established by
\cite{GuBenz1990} for a limited spectral range. This law predicts
that the duration of spikes at $f>$ 4.5 GHz should be less than 2
ms, which is well below the spectrometer temporal resolution, 5 ms.

\begin{figure}
\plotone{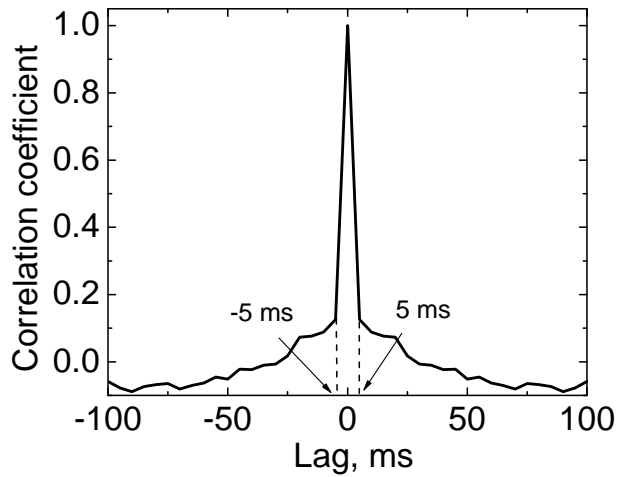} %{timecorr.eps}
\caption{Autocorrelation of the variable part of the signal averaged
over the frequency channels of the instrument. \label{fig:timecorr}}
\end{figure}

To check this prediction we studied the cross-correlation of the
recorded signal vs time lag and found that the adjacent time frames
are indeed entirely uncorrelated, Figure~\ref{fig:timecorr}. This
means that each spike does appear only once in the dynamic spectrum
in most of the cases. Therefore, the spikes are not resolved in time
by the PMO spectrometer. Thus, each time frame has to be processed
independently  from the adjacent time frames. This reduces a 2D
fitting problem to a sequence of 1D fitting problems, which is much
easier task.

\begin{figure}
 %\leavevmode
 %\centering\epsfxsize=200pt \epsfbox[65 400 630 800]
 \plotone{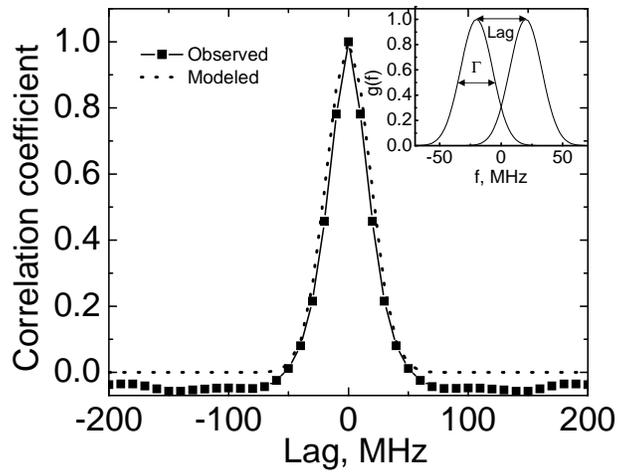} %{autocorr_freq.eps}
\caption{Autocorrelation of the instantaneous spectra in frequency
domain. The solid line shows the autocorrelation of the measured
signal. The squares correspond to the frequency lags equal to
integer multiples of the instrumental spectral resolution. The
dotted curve represents the best fit with the autocorrelation of a
purely gaussian signal $g(f)$ having the bandwidth $\Gamma$=30 MHz.
The inset shows two instances of the corresponding $g(f)$ shifted by
a certain lag. \label{fig:freqcorr}}
\end{figure}

On the contrary, the spike signal is well resolved in the spectral
domain. Figure~\ref{fig:freqcorr} shows autocorrelation of the
instantaneous spike signal in frequency domain. The autocorrelation
functions were calculated for each time frame having firm spike
signal. The obtained average autocorrelation function is shown in
Figure~\ref{fig:freqcorr} referenced as 'observed'. The squares
connected by the solid line show correlation coefficient calculated
at lag values of $n\times$10 MHz where integer $n$ denotes the
number of frequency channels of the spectrometer (having spectral
resolution of 10 MHz). Significant correlation between a few
adjacent channels clearly indicates that each spike is seen through
several spectral channels.

To understand the meaning of the shape and bandwidth of the
cross-correlation function obtained we performed a very simple
modeling as follows. We calculated the autocorrelation function of a
pure gaussian signal with a certain bandwidth. The corresponding
gaussian signal $g(f)$ is shown in the inset in
Figure~\ref{fig:freqcorr}. Remarkably, 'observed' curve can be
excellently fitted with the autocorrelation of a purely gaussian
signal of a certain width $\Gamma$ ('modeled' curve in
Figure~\ref{fig:freqcorr}). Even though, the similarity of the
'observed' curve to the 'modeled' curve can be a result of the
Central Limiting Theorem, rather than a similarity of individual
spike profiles to pure gaussian, it, nevertheless, implies the
presence of a characteristic bandwidth in the spike signal. The best
agreement between the 'modeled' and 'observed' plots is achieved for
$\Gamma=30$ MHz, which suggests that the bandwidth of the majority
of the spikes is around 30 MHz.

This value is not at all unexpected. \cite{CsillaghyBenz1993}
studied 196 individual spikes recorded in different events over
sufficiently broad spectral range to yield an empirical regression
law $\Gamma[MHz]=0.66f[MHz]^{0.42}$, which implies $\Gamma\approx
25$ MHz at the frequencies 4.5-6 GHz. Given a very large scatter of
individual measurements around this regression curve in
\citep{CsillaghyBenz1993}, we conclude that our finding of the
characteristic bandwidth around 30 MHz is in full agreement with the
results of \cite{CsillaghyBenz1993}. Dividing this value by the
typical frequency of $\approx 5$ GHz we get the relative bandwidth
of a spike $\Gamma_{rel}\approx 0.6\%$. Thus, we can identify this
value of $\Gamma_{rel}$ with a characteristic relative bandwidth of
spikes contained in the measured data.

\subsection{Spike Decomposition Technique}

\begin{figure}
%\leavevmode
 %\centering\epsfxsize=470pt \epsfbox[60 430 590 800]
  \epsscale{0.8}   \plotone{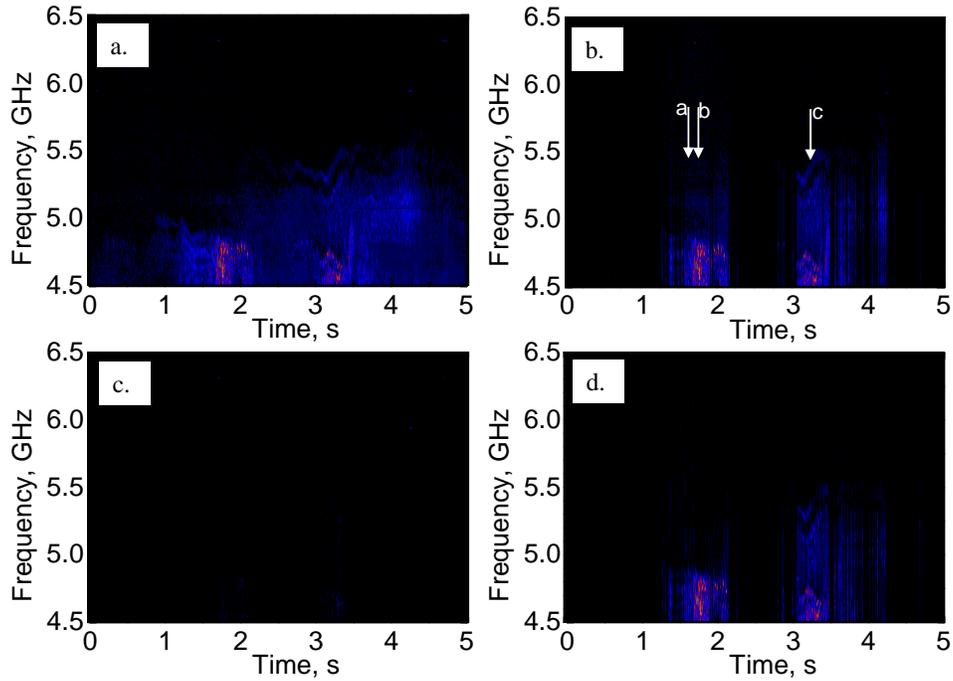} %{spectra.eps}
\caption{Dynamic spectra for the event under study: a -- calibrated
data from the PMO spectrometer, b -- the part of the spectrum
selected as input for spikes analysis, d -- fitted spectrum obtained
by the summation of the extracted Gaussian spikes superimposed on
the signal background level, c -- the absolute difference between
input spectrum (b) and fitted spectrum (d). The arrows in panel (b)
point out the time frames expanded in Figure~\ref{fig:fitex}. Time
is shown in seconds after 05:18:03.5 UT, April, 10, 2001.
\label{fig:spectra} }
\end{figure}

To handle with the data from the spectrometer we developed a new
technique utilizing a nonlinear multi-Gaussian spectral fit based on
"chi-squared" criteria to extract individual spikes from the
originally recorded spectra. %At the first stage we get rid of the
%frames where the intensity of the signal cannot be well
%distinguished from the noise for the whole frequency range so that
%only spectra containing unambiguous spiky signal were analyzed. This
%remained part of the overall dynamic spectra used for further
%analysis is shown in Figure~\ref{fig:spectra} (b).
%%\IR Not quite satisfied with the phrase. Suggest:
At the first stage we reveal the time frames where the intensity of
the signal cannot be well distinguished from the noise in the whole
frequency range. Such frames are excluded from the further analysis.
The retained part of the  dynamic spectra with distinguishable spiky
signal is shown in Figure~\ref{fig:spectra} (b). The technique
described below is sequentially applied to the set of retained time
frames. Remind that each instantaneous spectrum is treated as a set
of unique independent spikes.
%\IR suggest to omit the phrase above - it does not add any new information to what is told below
The input for the elementary fitting procedure is, therefore, an
instantaneous spectrum $S(f_j)$. The signal $S(f_j)$ is fitted with
a sum of model spikes superposed on a zero level $z$, which does not
depend on frequency but may change from frame to frame. Each spike
is assumed to have a Gaussian shape described by three parameters,
namely amplitude ($A_i$), mean ($f_{0i}$), and standard deviation
($\gamma_i$).
\begin{equation}
S^* \left( f \right) =z+ \sum\limits_{i = 1}^M {s_i \left( f
\right)},
\end{equation}
\begin{equation}
s_i \left( f \right) = s_i \left( {f,A_i ,f_{0i} ,\gamma _i }
\right) = A_i e^{ - \frac{{\left( {f - f_{0i} } \right)^2
}}{{2\gamma _i ^2 }}}.
\end{equation}
No frequency-dependent background component besides the statistical
noise is assumed. We shall often use the bandwidth $\Gamma_i$ (full
width at half maximum) to describe a spike rather than standard
deviation, both values related through
$\Gamma_i=2\sqrt{2\ln{2}}\gamma_i \approx 2.35\gamma_i$. It must be
noted that the bandwidths derived from the fitting (both $\Gamma_i$
and $\gamma_i$) are arbitrary real numbers rather than integer
multiples of the instrument spectral resolution of 10 MHz.

The fitting procedure starts from "guess" set of M spikes
\{$s_{0i}$\}, %their number and amplitudes being the same as of local
%maxima found in the signal above some threshold level, which was
%varied in various runs of the fitting to check the consistency and
%stability of the fitting results.
%\IR suggest
their number and amplitudes being that of local maxima found in the
signal above certain noise-threshold level. We changed this
noise-threshold level  in various runs of the fitting to check the
consistency and stability of the fitting results. The constrained
nonlinear minimization is performed with regard to the "chi-squared"
statistics
\begin{equation}
\chi ^2  = \sum\limits_j {\frac{{\left( {S\left( {f_j } \right) -
S^* \left( {f_j } \right)} \right)^2 }}{{\sigma _j ^2 }}}.
\end{equation}
using IDL routine CONSTRAINED$_{-}$MIN. The residual of this fitting
consists generally of the statistical noise (which is presumably
known based on the instrument characteristics) and a contribution
from numerous weak unresolved spikes, whose contribution is unknown.
Thus, the full uncertainties associated with a set of measurements
are unknown in advance.

\begin{figure}
% \leavevmode
 %\centering\epsfxsize=180pt \epsfbox[100 300 340 800]
 \epsscale{0.5} \plotone{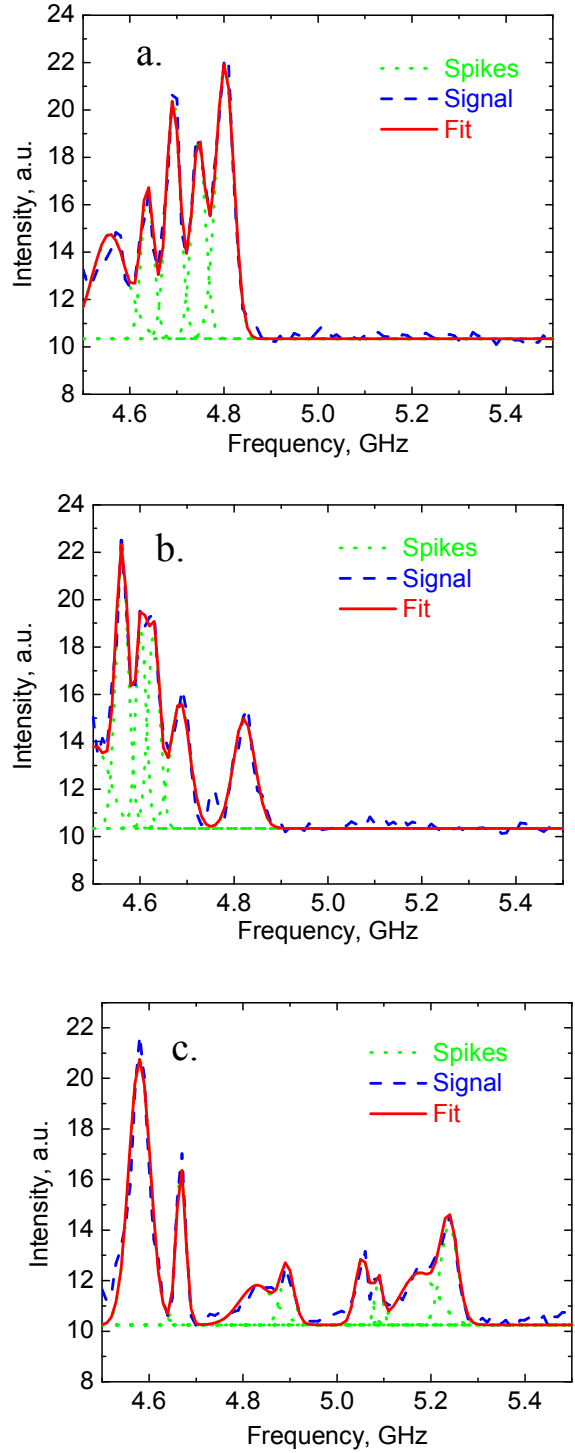} %{fitex1.eps}
\caption{Fit examples for three time frames marked by the arrows in
Figure~\ref{fig:spectra} (b). Dashed lines represent the input
spectra, solid lines are the sums of the extracted spikes. The
spikes are shown by dotted line \label{fig:fitex}}
\end{figure}

However, if we assume that all measurements have the same standard
deviation, $\sigma$, and that the model does fit well the data so
that $\chi^2\approx N-3M$, $N$ is the number of the data points used
for the fitting, then we can estimate $\sigma$ by first assigning an
arbitrary constant $\sigma_0$ to all points, next fitting for the
model parameters by minimizing $\chi^2$, and finally adopting
\begin{equation}
\sigma ^2  = \sum\limits_j {\frac{{\left( {S\left( {f_j } \right) -
S^* \left( {f_j } \right)} \right)^2 }}{{N - 3M}}}.
\end{equation}
After the minimization procedure is completed, each extracted spike
is tested for the statistical significance as follows. The spike is
excluded from the fitting function so that
\begin{equation}
S^{**} \left( f \right) = z + \sum\limits_{i = 1}^{M - 1} {s_i
\left( f \right)}.
\end{equation}
New minimization is performed and the new $\chi'^{'2}$ statistics is
calculated keeping the obtained $\sigma$. %The obtained $\chi'^{'2}$
%is compared to the previous $\chi^2$ .
%\IR suggest to omit the sentence above - has no info
The spike is excluded if this procedure does not decrease the
"goodness of fit" expressed by $\chi^2$ :
\begin{equation}
\frac{{\chi '^2 }}{{N - 3(M - 1)}} \le \frac{{\chi ^2 }}{{N - 3M}}.
\end{equation}
Figure~\ref{fig:fitex} shows an example of the fit for three
instantaneous spectra taken at the time frames pointed by the arrows
in Figure~\ref{fig:spectra}(b). The figures show very good agreement
between the initial raw signal and the fit spectrum being the sum of
the spikes extracted. Figure~\ref{fig:fitex}(b) also shows that
sometimes the fitting procedure may drop away a spike with a
moderate intensity still well pronounced. While such cases are
statistically rare, nevertheless we have performed a number of
spikes extraction procedures varying the parameters of the
constrained minimization as well as the value of the threshold used
to get Figure~\ref{fig:spectra}(b) from Figure~\ref{fig:spectra}(a).
The corresponding variation of the output statistical parameters of
the extracted spike ensembles allowed us to estimate the statistical
errors as well as the mean quantities given in the next subsection.

\subsection{Spikes Extraction Results}

\begin{figure}[ht]
% \leavevmode
 %\centering\epsfxsize=200pt \epsfbox[100 370 670 800]
  \plotone{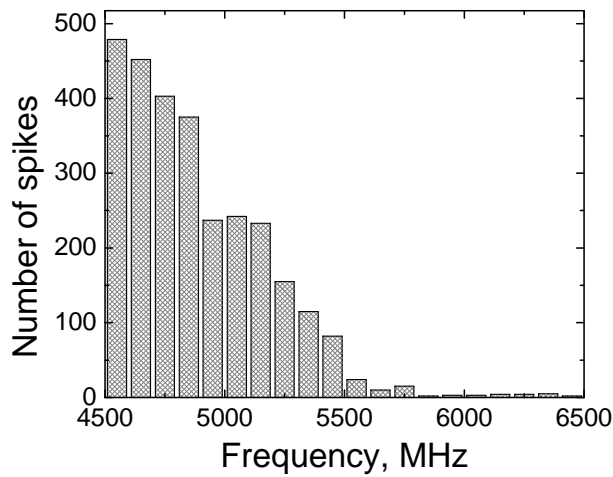} %{frequencyDistribution.eps}
\caption{Distribution of the mean frequencies of the extracted
spikes. \label{fig:frequencyDistribution}}
\end{figure}

\begin{figure}[ht]
% \leavevmode
 %\centering\epsfxsize=200pt \epsfbox[100 370 670 800]
  \plotone{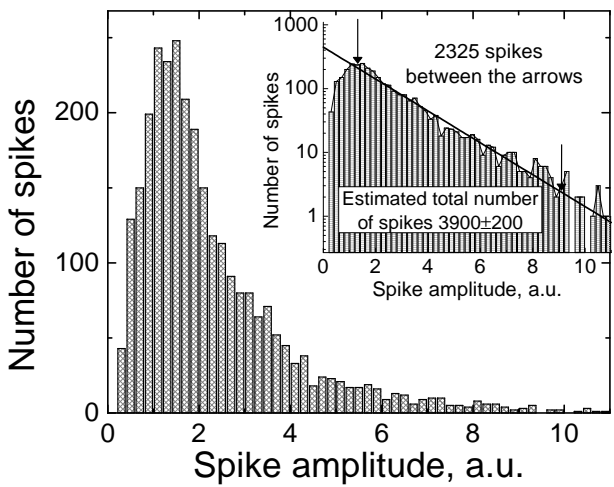} %{amplitudeDistribution.eps}
\caption{Amplitude distribution of the extracted spikes. The inset
shows the same distribution in log-linear scale. The solid line is
the approximation of the decreasing part of the plot containing 2325
spikes by an exponential distribution. Extrapolation of this line
towards zero amplitude allows estimating the total number of the
spikes in the time frames selected for the analysis as about 4000.
\label{fig:amplitudeDistribution}}
\end{figure}

\begin{figure}[ht]
% \leavevmode
 %\centering\epsfxsize=400pt \epsfbox[80 580 650 800]
  \epsscale{1} \plotone{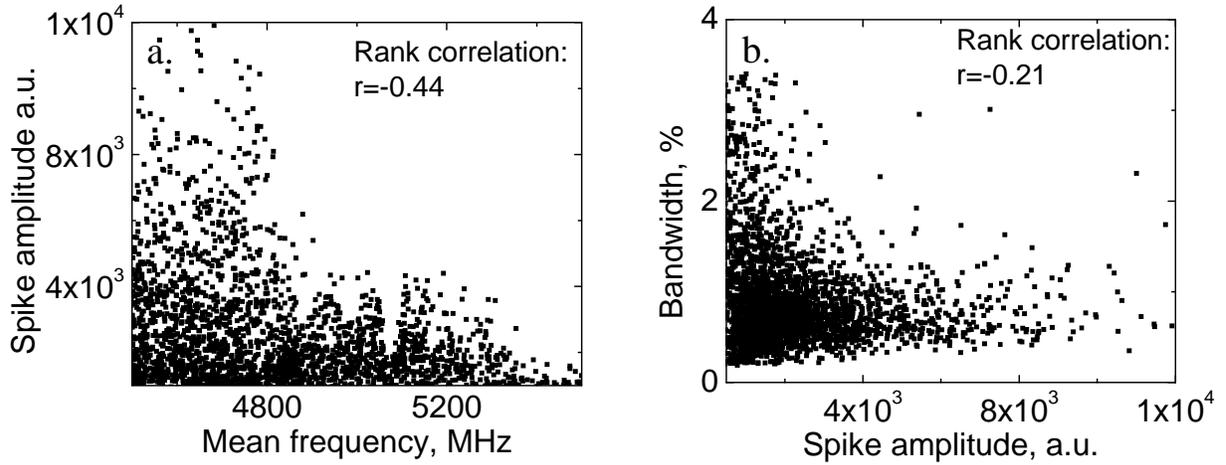} %{correlations.eps}
\caption{Correlation between mean frequency and spike amplitude
(left) and between spike amplitude and relative bandwidth (right).
These plots show that lower-frequency spikes are typically stronger,
and stronger spikes tend to be more narrowband,  though the
correlation coefficients are somewhat low.} \label{fig:correlations}
\end{figure}

\begin{figure}[ht]
% \leavevmode
 %\centering\epsfxsize=200pt \epsfbox[100 370 670 800]
  \epsscale{0.5} \plotone{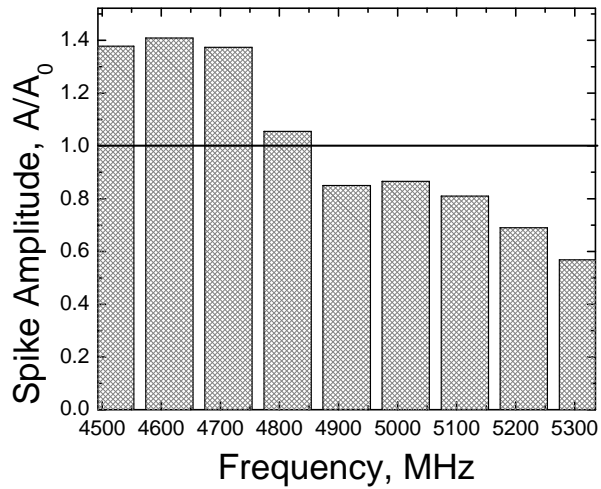} %{amplitudeMoving.eps}
\caption{Moving average of spikes amplitudes vs. spike mean
frequency. At least 100 spikes were averaged for each frequency
range. \label{fig:amplitudeMoving}}
\end{figure}

The developed technique allowed us to extract about 3000 spikes.
Each spike $i$ is characterized by its amplitude $A_i$, mean
frequency $f_{0i}$ and the bandwidth $\Gamma_i$. The distribution of
the spike mean frequencies  is shown in
Figure~\ref{fig:frequencyDistribution}. In obvious agreement with
the raw data spectra (Figure~\ref{fig:spectra} (a),(b)) the number
of spikes decreases with the frequency increase. A small deviation
from this tendency can be seen near $\approx5$ GHz. This corresponds
to the sharp end of the first subcluster at about 4.9 GHz and also
to the malfunction of a few frequency channels of the instrument
clearly seen as a blue horizontal stripe at $\approx 5.1$ GHz in
Figure~\ref{fig:spectra} (a). The amplitude distribution of the
spikes is shown in Figure~\ref{fig:amplitudeDistribution}. Note that
only the decreasing part of the distribution has sense while the
initial increasing part is possibly an artefact caused by skipping
the spikes which undergo the average noise level. The decreasing
part of the amplitude distribution follows an exponential law (see
the inset in Figure~\ref{fig:amplitudeDistribution}), which allows
estimation of the number of weak missing spikes to be about 30$\%$
of the extracted spikes. The spike amplitude correlates with its
frequency as Figure~ \ref{fig:correlations} shows, which means that
both more spikes and stronger spikes are produced at lower
frequencies; the calculated rank correlation coefficient is
$r=-0.44$. This tendency is  also apparent from
Figure~\ref{fig:amplitudeMoving}, which presents the moving average
for the spike amplitude versus spike central frequency.

A striking finding of the presented spike analysis is distribution
of the spike relative  bandwidths. This commonly used dimensionless
parameter is defined as spike width at half-maximum divided by the
central frequency of the spike. We checked and found that the
bandwidth distribution has no imprint of the discretness of the
spectral channels: integer multiples of 10 MHz do not display any
local peak compared with neighboring non-integer values. All values
for the relative bandwidth present in the histogram are in excess
the minimum spike bandwidths reported in the literature for other
events \citep{MessmerBenz2000, Wang_etal_2003}.

\begin{figure}[ht]
% \leavevmode
 %\centering\epsfxsize=200pt \epsfbox[100 370 670 800]
  \plotone{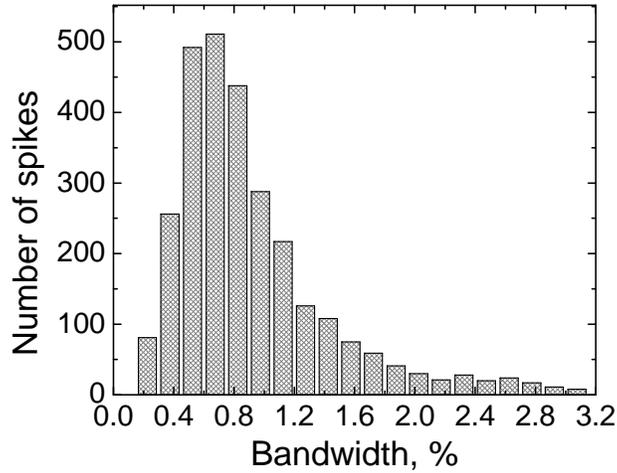} %{bandwidthDistribution.eps}
\caption{Bandwidth distribution of the extracted spikes. The binsize
is taken to be 0.15\% to avoid any coincidence with the spectral
resolution of the instrument. Note prominent asymmetry of the
distribution. \label{fig:bandwidthDistribution}}
\end{figure}

The distribution of the spike relative bandwidths appears to be
remarkably asymmetric. It has a skew shape with rapid increase at
low values of the relative bandwidth followed by maximum at 0.6\%
and smooth tail approaching zero at approximately 3\%
(Figure~\ref{fig:bandwidthDistribution}). Note that the peak of the
distribution (0.6\%) corresponds to the characteristic spike
bandwidth found in \S 2.2 from the correlation analysis.

The asymmetry of the distribution can be characterized by the
skewness which is the third moment:
\begin{equation}
S = \frac{{\left\langle {\left( {f - f_0 } \right)^3 } \right\rangle
}}{{\sigma ^3 }}.
\end{equation}
For the spike ensemble under study the overall skewness is about
1.6. The deviation from the normal distribution is estimated by the
forth moment, kurtosis, defined as
\begin{equation}
 K = \frac{{\left\langle {\left( {f - f_0 }
\right)^4 } \right\rangle }}{{\sigma ^4 }},
\end{equation}
which is $K\approx6$ for our case, in contrast to that of the normal
distribution $K_{norm}=3$. Rather weak correlation is found between
amplitude and spike relative bandwidth,
Figure~\ref{fig:correlations}, the appropriate rank correlation
coefficient is only $r\approx-0.2$. No correlation is observed
between mean frequency and relative bandwidth of the spikes (the
appropriate rank correlation coefficient is $r\approx0.04$).

\begin{figure}[ht]
%\leavevmode
 %\centering\epsfxsize=200pt \epsfbox[100 370 670 800]
  \plotone{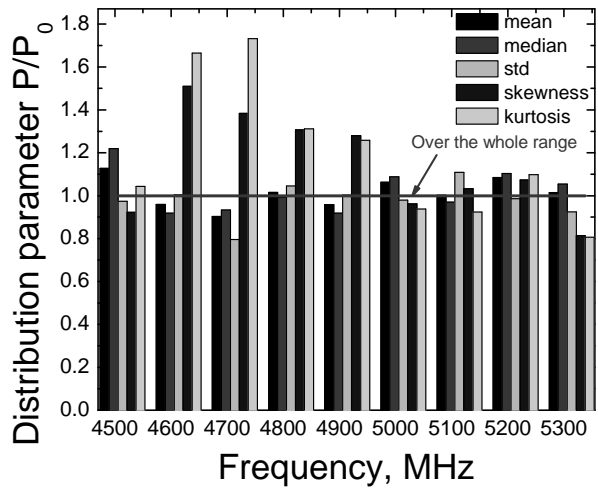} %{distributionParametersmoving.eps}
\caption{Parameters of the spikes bandwidth distribution versus
spike mean frequency. For each frequency range the distributions
were calculated over at least 100 spikes.
\label{fig:distributionParametersmoving} }
\end{figure}

The large number of individual spikes identified allows for analysis
of the spike properties in restricted spectral regions. Accordingly,
we looked for a dependence of the bandwidth distribution shape on
the spike central frequency. However, no unambiguous  trends have
been found. Figure~\ref{fig:distributionParametersmoving} displays
the frequency dependence of the moments of the distribution, such as
mean, median, standard deviation, skewness, and kurtosis, normalized
to the corresponding global values gathered in Table~1. %are plotted as moving averages
%versus spike central frequency.
The figure shows no significant deviation of any of these parameters
$P$ from its overall average value $P_0$.

\section{Local Trap Model}

\begin{figure}[ht]
%\keepaspectratio
    \epsscale{0.5}
  \plotone{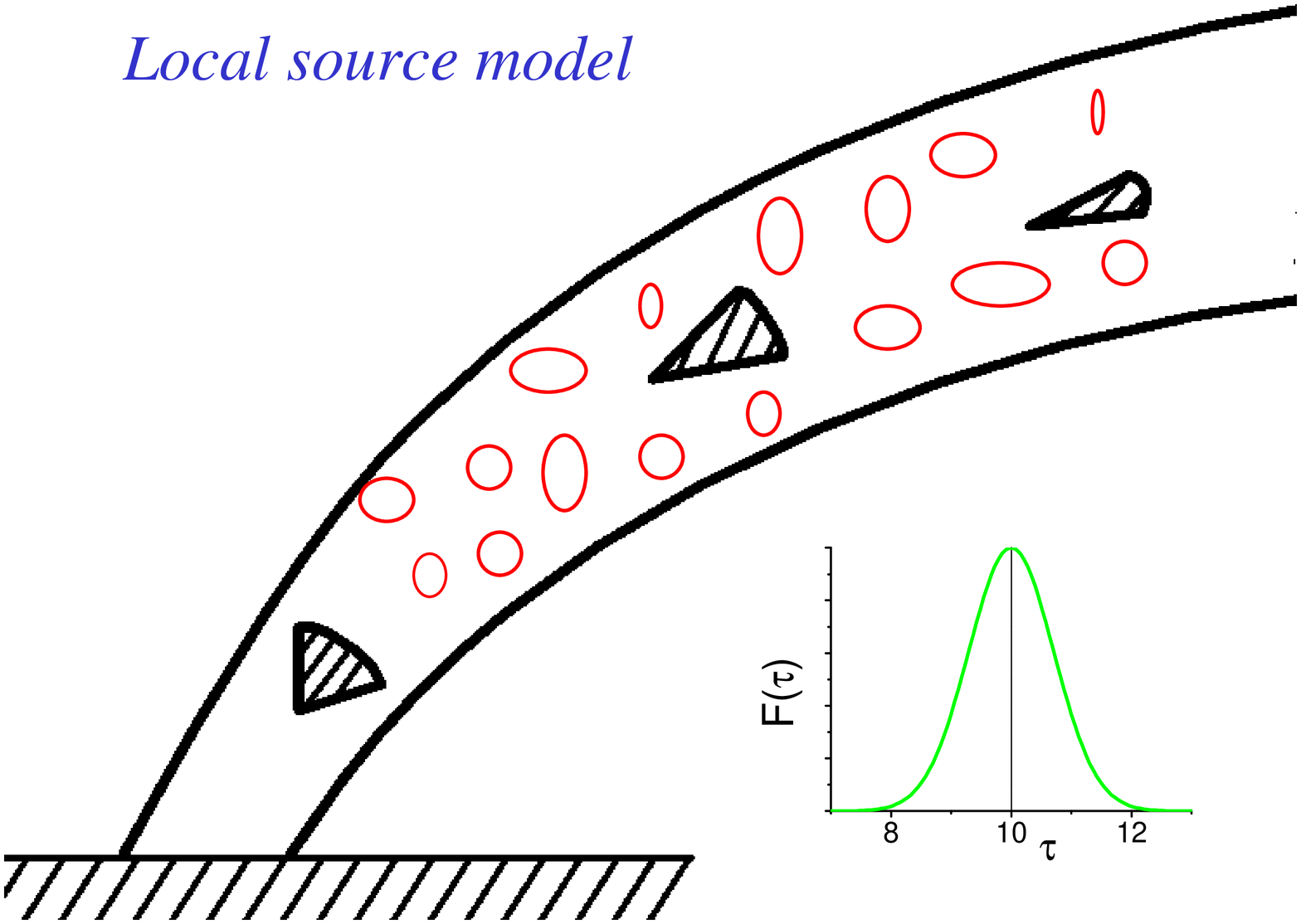} %{local_traps.eps}
\caption{Sketch for the local trap model of spike generation. Red
ovals show local traps where the pitch-angle anisotropy of the fast
electron distribution is enhanced due to fluctuations of the
magnetic field provided by magnetic turbulence in the large-scale
trap. The inset displays the adopted symmetric (gaussian)
distribution of the local spike sources over the ECM optical depth
$\tau$ with the mean value $\tau=10$ and dispersion $\sigma=1$.
\label{local_traps}}
\end{figure}

As has already been  noted most of the spike properties are
consistent with a local trap model \citep{FlMel1998,FlGaryNita2003}.
This model adopts that a spike cluster is produced at a significant
portion of a magnetic trap (Figure~\ref{local_traps}), where a
loss-cone distribution of the trapped fast particles is formed due
to emptying the loss cone as a result of the electron precipitation
into the foot points. The overall pitch-angle anisotropy is moderate
on average provided that the mean fast electron distribution is at
about the marginal stability state in respect to ECM generation. An
important ingredient of the local trap model is a magnetic
turbulence, which gives rise to local variation of the fast electron
distribution anisotropy. Under favorable conditions, this turbulence
will increase the anisotropy to the extent sufficient for the ECM
instability to develop at some local places inside the large-scale
magnetic trap. Such favorable places represent those local spike
sources, quasi-randomly distributed over the trap.

As a byproduct of the key role of the magnetic turbulence in forming
the spike local sources, the model suggests that the small-scale end
of the magnetic turbulence spectrum persists in each local spike
source. Therefore,  the spikes are formed in a source, where random
magnetic inhomogeneities are superimposed on the mean magnetic field
of the source. \cite{Fl_2004b} developed a theory of  the  ECM
spectral bandwidth taking into account the  stochastic
irregularities of the magnetic field at the source. He obtained that
the bandwidth of a single ECM peak generated in such a source is
specified by the relation
\begin{equation}
\Gamma  = \Gamma _0 e^{\frac{a}{{2\Gamma _0^2 }}},
\end{equation}
where  $\Gamma_0$  is the "natural" bandwidth of the ECM peak in the
uniform source with the optical depth $\tau$ in the peak maximum,
\begin{equation}
a = \frac{{s_0^2 }}{2}\frac{{\left\langle {\delta B^2 }
\right\rangle }}{{B^2 }}
\end{equation}
is a "turbulence parameter", defined by the ECM harmonic number
$s_0$ and the magnetic turbulence energy density $\left\langle
{\delta B^2 } \right\rangle /8\pi $ normalized by the magnetic
energy density $B^2 /8\pi $. We note that the magnetic
inhomogeneities give rise to quite a strong ECM broadening when $a
\gsim \Gamma_0^2$. Since $\Gamma_0^2 \ll 1$, rather weak random
inhomogeneities of the magnetic field provide large ECM broadening.

To account for the observed spike bandwidth distribution we
developed a simple model based on the electron cyclotron maser (ECM)
emission within the local source model. Specifically, we make use
that the natural spike bandwidth is about $0.1-0.3\%$ in agreement
with calculations of the ECM natural spectral bandwidth
\citep{Fl_2004_AL}. We postulate a symmetric gaussian distribution
of the natural bandwidth over the spike local sources. Then, we
adopt that the turbulent parameter $a$ has another gaussian
distribution, not correlated with the natural bandwidth
distribution.

\begin{figure}[ht]
% \leavevmode
 %\centering\epsfxsize=200pt \epsfbox[100 370 670 800]
  \plotone{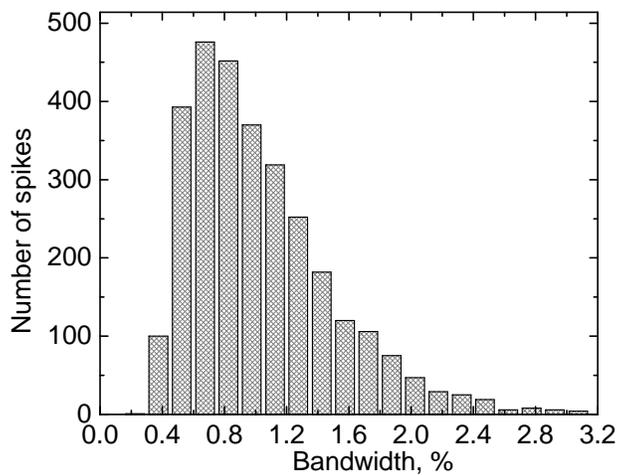} %{bandwidthmODEL.eps}
\caption{Spike bandwidth distribution obtained from the local trap
model. \label{fig:bandwidthmODEL}}
\end{figure}

\begin{figure}[ht]
% \leavevmode
 %\centering\epsfxsize=180pt \epsfbox[60 320 330 790]
  \plotone{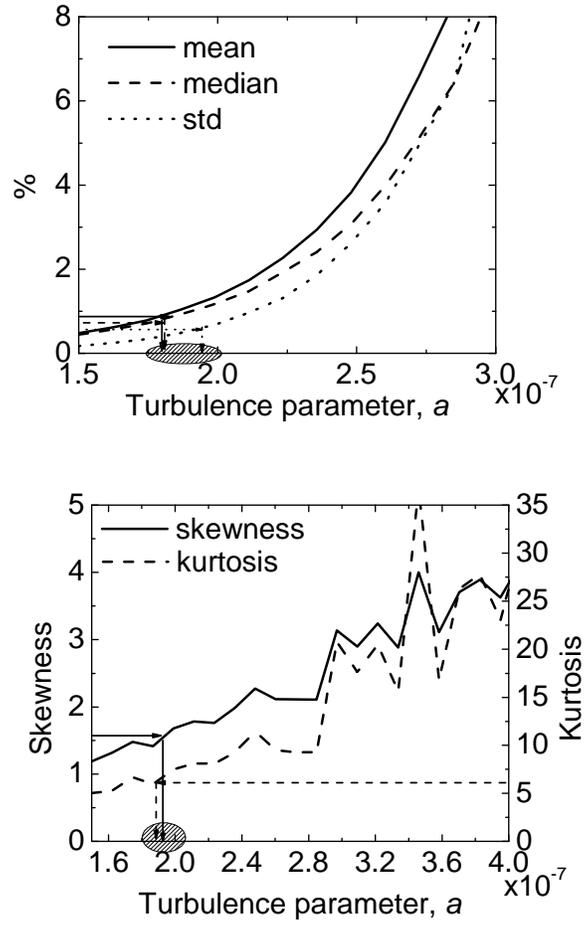} %{distributionParameters.eps}
\caption{Dependences of the spike bandwidth distribution parameters
on turbulence parameter $a$ obtained from the local trap model
described in the text.} \label{fig:distributionParameters}
\end{figure}

Varying the pairs of $\Gamma_0$ and $a$ within those two parent
distributions we are able to produce artificial sets of spikes. The
properties of these artificial spike distributions are specified by
the parameters of the adopted gaussian distributions, i.e., mean
values and dispersions of $\Gamma_0$ and $a$. In our modeling we
kept mean $\Gamma_0$ value constant at $0.2\%$ level, while varied
$a$ value to study the dependence of the distribution moments on
$a$. The dispersion of both values was taken to be about $15\%$. An
example of the distribution produced by the model is given in
Figure~\ref{fig:bandwidthmODEL}. Eventually, we generated 50 sets
with 3000 artificial spike in each set and calculated four first
moments of this distributions, which are plotted in
Figure~\ref{fig:distributionParameters}.

Figure~\ref{fig:distributionParameters} displays also values of the
observed distribution moments. Remarkably, all the observed moment
values are consistent with corresponding model values for $a \approx
2\cdot10^{-7}$, which is especially important because the model
includes only one free parameter to yield all four moments together
with the correct values. Saying in other words, the comparison of
the observed and model moments offers an elegant method of studying
the small-scale magnetic turbulence in the sources of spike
clusters. We note that the method is highly sensitive to the
magnetic irregularities and capable of detecting the turbulence with
remarkably low level $\left\langle {\delta B^2 } \right\rangle /B^2
\sim 10^{-7}$ or even less.

\section{Discussion}

Although the phenomenon of the narrowband spikes and related fine
spectral and temporal structures has not been well understood yet,
the permanently accumulated data and its detailed and critical
analysis in the context of the competing source models has given
rise to a significant progress. In particular, there are currently
strong evidences that narrowband millisecond spikes  appearing at
the main flare phase are  generated by ECM emission. It follows,
e.g., from detailed comparison between the observed properties of
solar radio spikes and predictions of various theoretical models
\citep{FlMel1998}. Even stronger evidence is found recently from the
analysis of correlations between the spikes and accompanying
microwave continuum \citep{FlGaryNita2003}: they found, that
spike-producing radio bursts reveal smaller plasma frequency to
gyrofrequency ratios than other radio bursts, and the strongest
averaged flux of the spike emission is observed when the fast
electrons display the hardest energy spectra and the most
anisotropic angular distribution.

However, it is still unclear if spikes in all events are produced by
the same emission mechanism, or there are different subclasses of
the spikes. Indeed, only a half of all spike clusters analyzed by
\cite{AschwGu1992} correlate well with simultaneous hard X-ray (HXR)
emission, the other half reveal weaker or no correlation. Then,
\cite{Benz_etal_2002} studied the location of narrowband spikes
occurred at frequencies $\lsim 450$ MHz at the flare decay phase.
The spike source was found to be located far away from the main
flare location, which was interpreted as a post-flare electron
acceleration high in the corona. Such spike bursts are likely
different in nature from the spike clusters originating at the
impulsive and main flare phase, which are highly correlated with HXR
and microwave flare emission. We believe that more progress in
understanding the nature of the spikes can be expected when
broadband imaging radio instruments start to operate.

Similar dynamic spectra of radio emission produced by different
mechanisms can be understood if the (quasi-random) distribution of
the spikes in frequency and time is related to the global source
structure rather than microscopic emission process. For example,
\cite{BaKar2001} developed a turbulent model of the spike source,
which can result in dynamic spectra typical for spikes for different
emission mechanisms. Therefore, detailed case studies as well as
statistical studies capable of distinguishing between competing
emission mechanisms are exceedingly important.

This paper reports a spike cluster at high frequencies, 4.5--6 GHz,
which is relatively rare type of events. No statistics of spikes has
been available at this spectral range. To remedy this situation we
developed a method capable of extracting spikes from dense clusters
of overlapping spikes. Applying this method we extracted a few
thousand spikes sufficient for a detailed analysis of spike
properties. In particular, besides finding the mean and the variance
of the distribution of the spike relative bandwidth, we were able to
confidently determine higher moments, namely skewness and kurtosis.
We found this distribution to be highly asymmetric and to deviate
confidently from the normal distribution. Although similar
asymmetric distributions at lower frequencies have already been
reported based on manual selections of well isolated gaussian spikes
in less dense regions of the spike clusters
\citep{CsillaghyBenz1993,MessmerBenz2000}, those previous studies
involved much lower numbers of manually selected spikes, therefore,
the statistics was less significant.

Complimentary, we developed a simplified theoretical model of spike
cluster generation in a magnetic trap with random inhomogeneities
based on the renormalized theory of the ECM spectral properties
taking into account these inhomogeneities. The simulated spike
distribution is remarkably similar to the observed one for this
particular event, which is a strong evidence in favor of the ECM
mechanism of the spike generation within the local trap model with
the magnetic turbulence.

Thus, the demonstrated agreement between the model and observations
%represents a new strong evidence in favor of ECM mechanism of the
%spike generation and
may suggest a new efficient tool of measuring weak magnetic
inhomogeneities in the spike sources. Since the turbulent parameter
$a$ depends on the emission harmonics number, it is highly desirable
to determine at what harmonics of the gyrofrequency the spikes are
produced. A direct answer to this question could be obtained from
imaging observations of the spike source, which are currently
unavailable at the considered frequency range. Therefore, we have to
use some indirect approach.

The ECM emission requires both nonthermal electrons and sufficiently
strong magnetic field to coexist at the spike source.
Figure~\ref{fig:MDI} displays the photospheric magnetic field
regions corresponding to fundamental, second, and third
gyroharmonics falling to the 4.5--5.6 GHz range and contours of the
optically thin microwave radiation at 17 GHz indicating the region
where the fast electrons are present. Although the spikes are
produced in the coronal rather than photospheric sources, we
believe, that the spatial locations of strong coronal magnetic
fields are correlated with regions of strong photospheric magnetic
fields. Thus, the presence of strong photospheric magnetic field is
a necessary condition for the presence of strong coronal field
nearby. Inspection of this figure shows that generation of the
fundamental ordinary mode ECM emission is almost certainly excluded,
while the fundamental extraordinary ECM emission is not too probable
even though it cannot be confidently excluded.

Apparently, the most probable location of the spike source is the
position of  the peak of the positive Stocks V distribution
overlapping with a tongue of the Stocks I distribution occurring on
top of the strong positive magnetic field region, because high
circular polarization ($\sim50\%$) of the microwave continuum
radiation is a strong indication in favor of large magnetic field in
this part of the radio source \citep{BBG}. Thus, the second
extraordinary harmonics is the most probable emission mode in the
spike cluster studied, although the third extraordinary and the
second ordinary harmonics cannot be firmly excluded. In any case,
the performed study involving the forward modeling offers new
sensitive tool of detecting and measuring the magnetic turbulence in
the spike sources.

\section{Conclusions}

We report a high-frequency dense cluster consisting of many
overlapping spikes and suggest a tool of decomposing the cluster
onto individual gaussian spikes. The large number of the identified
spikes allows for much more detailed statistical analysis of them,
than has ever been published. Then, we develop a theoretical model
of the spikes produced by ECM mechanism in numerous local sources
with random magnetic inhomogeneities. We find that the model and the
observed distributions of the spike relative bandwidth agree
excellently with each other, therefore, the presented observations
support the adopted local source model with magnetic turbulence.

Since a specific shape of the model distribution is highly sensitive
to the magnetic turbulence, this study suggests an elegant  and
straightforward way of studying magnetic turbulence in the spike
sources. In favorable conditions this study can be performed in the
spectral and temporal domains giving rise to the information of the
spatial distribution and evolution of the magnetic turbulence. We
note that the level of the magnetic turbulence in the considered
event is very weak, $a<10^{-6}$, and most probably not enough for
the bulk particle acceleration in the spike source. We interpret
this result as an important indication that the spikes are a
secondary phenomenon, rather than a manifestation of the primary
energy release and particle acceleration.

The developed tool, while powerful and promising,  requires further
analysis and development. In particular, we need detailed modeling
to understand better its capability and limits of applicability to
denser spike clusters. This is especially important for studying
long-lasting clusters with the spike density strongly changing in
time. Then, analysis of larger number of the spike clusters
throughout the whole spectral range where the spikes are observed
will be necessary to convert this tool into a kind of routine flare
diagnostics.

%A highly sensitive method of small-scale magnetic turbulence is
%suggested
%
%More events is needed.
%
%Limits of its applicability to more dense clusters are unclear.
%Modeling is needed.
%
%Does this hold for other events?

\acknowledgments %The authors express appreciation to Prof. Kiyoto
%Shibasaki and Prof. Hiroshi Nakajima  for the NoRP calibration
%corrections they provided.
This work was supported in part by NSF grants ATM-0607544 and
ATM-0707319  to New Jersey Institute of Technology,  by the Russian
Foundation for Basic Research, grants No. 06-02-16295, 06-02-16859,
06-02-39029,  NFSC projects No. 10333030, 10773032, and ''973''
program with No. 2006CB806302. We have made use of NASA's
Astrophysics Data System Abstract Service, SOHO/MDI data as well as
radio data from Nobeyama Radioheliograph and Polarimeters.

\clearpage

\begin{deluxetable}{l l@{~~~} l} %{l}
\
%PP
\tablewidth{3.3in} \tablecaption{Bandwidth distribution parameters
\label{table_1}} %{r}
%\label{tspix}
%\tabletypesize{small}
\tablehead{ \colhead{Moment of Distribution} & & \colhead{value}  }
\startdata
%  name    altname     ar                  ax
%                                * here means ar uncertainty estimated
Mean ($f_0$), \% & & 0.91$\pm$0.02  \\

Median, \% & & 0.77$\pm$0.01  \\
%  ar GreenCat   ax MathesonS2005, avg. for plerion,uncert. reflects range
Standard deviation ($\sigma$),  \% &   & 0.55$\pm$0.04 \\
% ar BockG2005, uncert estimated; ar Gotthelf2003
Skewness, S     &     & 1.6$\pm$0.2  \\
% ar VelusamyB1988; uncert estimated  ax from Gotthelf2003 (also a value of -0.88 from Lu+2002),
 Kurtosis, K  &     &  5.9$\pm0.6$  \\
\enddata

\end{deluxetable}

\clearpage

\end{document}